\documentclass[conference]{IEEEtran}
\IEEEoverridecommandlockouts
\usepackage{cite}
\usepackage{amsmath,amssymb,amsfonts}
\usepackage{graphicx}
\usepackage{textcomp}
\usepackage{xcolor}
\def\BibTeX{{\rm B\kern-.05em{\sc i\kern-.025em b}\kern-.08em
    T\kern-.1667em\lower.7ex\hbox{E}\kern-.125emX}}

\usepackage[margin=1.8cm]{geometry}
\usepackage{siunitx}
\usepackage{mathtools}
\usepackage{subcaption}
\usepackage{algpseudocode}
\usepackage{algorithm}
\usepackage{threeparttable}
\usepackage{bm}
\usepackage{url}
\usepackage{tabu}
\usepackage{multirow}
\usepackage{pgfplots}

\begin{document}

\title{A Power-Efficient Binary-Weight Spiking Neural Network Architecture for Real-Time Object Classification}

\author{\IEEEauthorblockN{Pai-Yu Tan$^1$, Po-Yao Chuang$^1$, Yen-Ting Lin$^1$, Cheng-Wen Wu$^{1, 2}$, and Juin-Ming Lu$^{1, 3}$}
  \IEEEauthorblockA{
	\textit{$^1$Department of Electrical Engineering, National Tsing Hua University, Hsinchu, Taiwan} \\
	\textit{$^2$Department of Electrical Engineering, National Cheng Kung University, Tainan, Taiwan} \\
	\textit{$^3$Infor. and Comm. Res. Labs, Industrial Technology Research Institute, Hsinchu, Taiwan}
  }
}

\maketitle
\bibliographystyle{ieeetr}

\begin{abstract}
Neural network hardware is considered an essential part of future edge devices. In this paper, we propose a binary-weight spiking neural network (BW-SNN) hardware architecture for low-power real-time object classification on edge platforms. This design stores a full neural network on-chip, and hence requires no off-chip bandwidth. The proposed systolic array maximizes data reuse for a typical convolutional layer. A 5-layer convolutional BW-SNN hardware is implemented in 90nm CMOS. Compared with state-of-the-art designs, the area cost and energy per classification are reduced by 7$\times$ and 23$\times$, respectively, while also achieving a higher accuracy on the MNIST benchmark. This is also a pioneering SNN hardware architecture that supports advanced CNN architectures.
\end{abstract}


\section{Introduction}
In recent years, extensive research on Convolutional Neural Network (CNN) has shown breakthroughs in many computer vision tasks, such as image classification, face recognition, and object detection. However, the substantially high computation complexity makes it extremely power-hungry. Because of the energy and latency issues, there is an emerging need of processing CNN directly on end-point devices. Most current end-point devices lack the computing capability for real-time processing of CNN. Therefore, this has triggered research that focuses on hardware and software architectures for low-power CNN computation on end-point and edge platforms.

\begin{figure}[h]
	\centering
	\includegraphics[width=\linewidth]{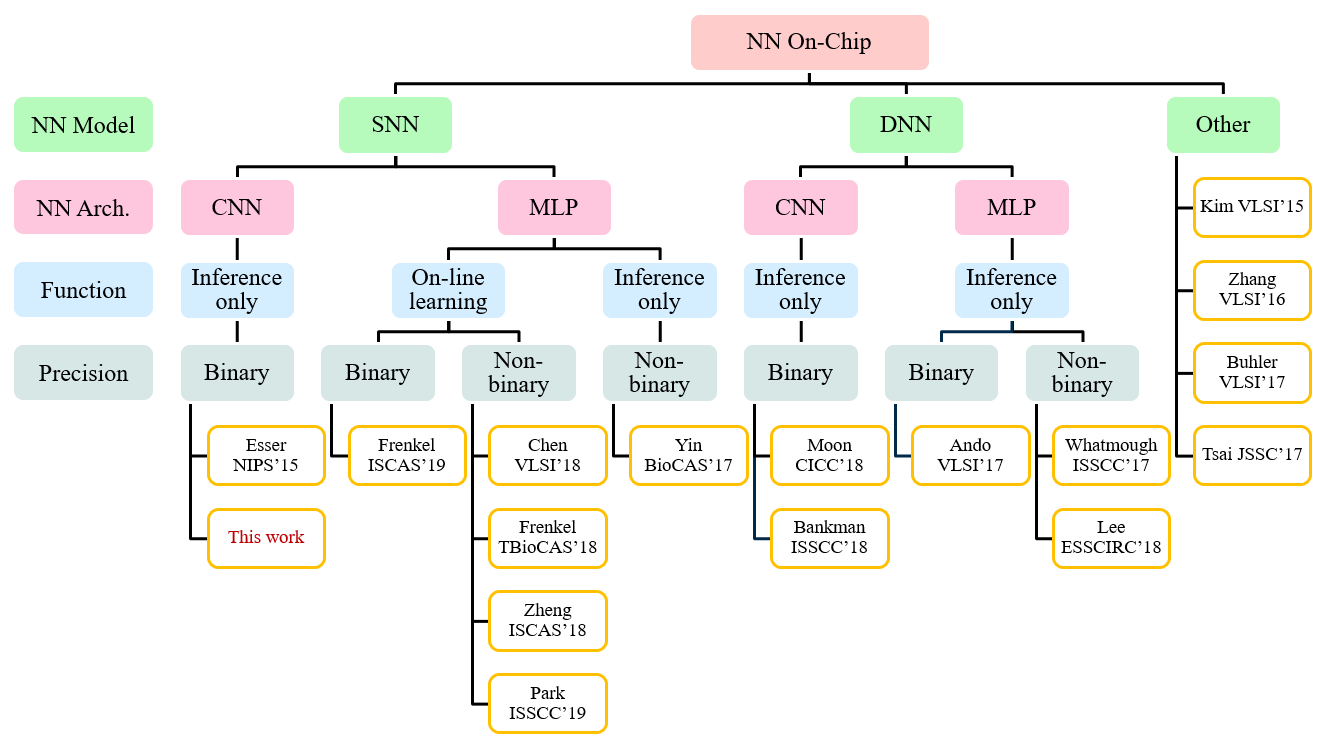}
	\caption{NN on-chip designs \cite{frenkel201965, lee2018wide, moons2018binareye, frenkel20180, zheng2018low, whatmough201714, park20197, yin2017algorithm, buhler20173, tsai201741, kim2015640m, esser2015backpropagation, essera2016convolutional, chen2019a4096neuron, bankman2018always, ando2017brein, zhang2016machine}.}
	\label{fig:prev_design}
\end{figure}

Some recent NN on-chip designs are summarized and shown in Fig. \ref{fig:prev_design}. DNN inference on end-point devices is promising. However, on-line training of DNN generally requires much higher computing power. Embedded on-line learning is considered to be a key feature for future edge devices, as it enables on-the-fly adaptation to the environment. As a result, a bio-inspired spiking neural network (SNN) is currently explored as an alternative neural network model because its spike-based learning rule is relatively simple to implement in hardware, e.g., \cite{chen2019a4096neuron, frenkel20180, zheng2018low, park20197} have demonstrated some on-line learning hardware for hand-written digit recognition. However, these designs only demonstrate their inference and learning on simple network structures, such as MLP. Little attention has been paid to support advanced network architectures, such as CNN, DenseNet \cite{huang2017densely}, Inception \cite{szegedy2016rethinking}, and MobileNet \cite{howard2017mobilenets}, which are more appropriate for real applications. To date, only IBM TrueNorth \cite{essera2016convolutional} and Intel Loihi \cite{davies2018loihi} have the ability to handle these advanced network architectures. However, these two chips are digital processors that are intended for simulating different network topologies, hence are not optimized for low-power edge platforms. 

The primary aim of this paper is to explore a power-efficient SNN hardware architecture for advanced CNN structures. We use binary-weight SNN (BW-SNN) to simplify the operations and minimize the on-chip storage. In order to efficiently process the CNN, we properly map the entire convolutional layer to a 2D systolic array, which is the basic building block of our BW-SNN hardware architecture. We demonstrate this architecture with a 5-layer convolutional BW-SNN implemented in 90nm CMOS. The post-layout simulation results show that the area, energy, and accuracy are better than state-of-the-art designs on MNIST benchmark. In addition, our design supports advanced CNN topologies on SNN hardware.

\section{BW-SNN Hardware Architecture}
\label{sec:architecture}

\subsection{Overall Architecture}
\begin{figure}[h]
	\centering
	\includegraphics[width=\linewidth]{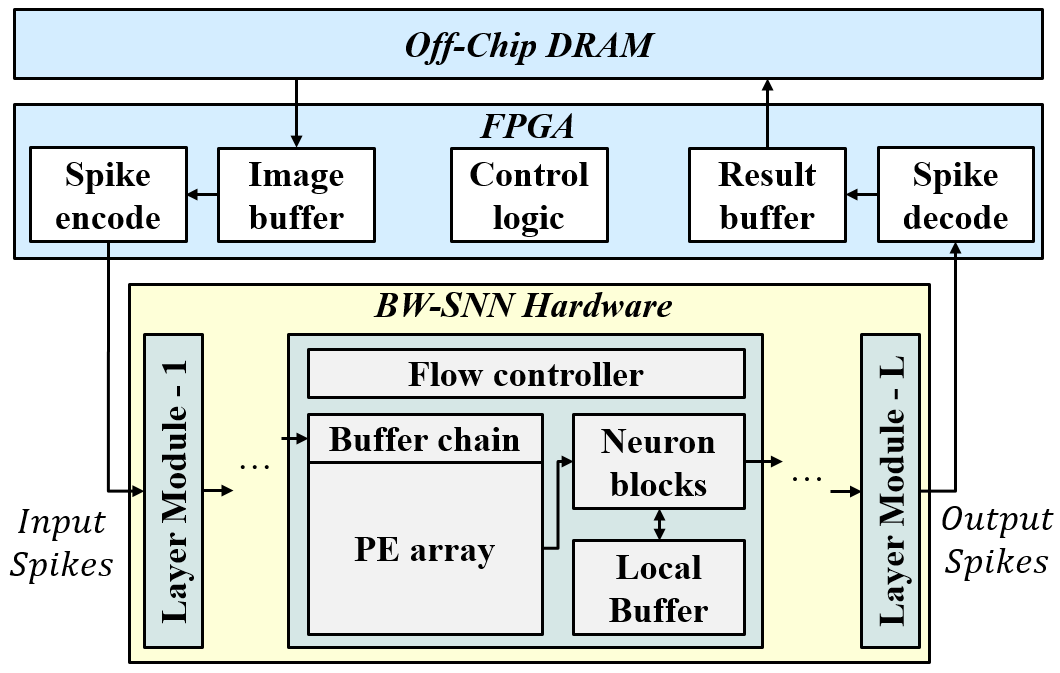}
	\caption{Binary-weight SNN hardware architecture.}
	\label{fig:hw_architecture}
\end{figure}

Figure \ref{fig:hw_architecture} shows the proposed BW-SNN hardware architecture. This architecture is designed based on the network topology of a BW-SNN, so the functionality of each layer is a specialized Layer Module that is the basic building block of the network. The inputs of a \textit{layer module} are the spikes from the previous layer module. These spikes flow into the systolic array consisting of a buffer chain and a PE array, which will be discussed in Sec. \ref{sec:systolic}. This systolic array computes weight sums for the subsequent \textit{neuron blocks}. The neuron block is a digital circuit that implements the \textit{Integrate-and-Fire} (IF) neuron function. The neuron blocks access the IF neuron parameters, such as membrane potentials, firing thresholds, and biases, from the local buffer, then update the neuron states and generate output spikes for the next layer module. 

\subsection{Data Flow and Systolic Array Architecture}
\label{sec:systolic}

\begin{figure}[h]
	\centering
	\includegraphics[width=\linewidth]{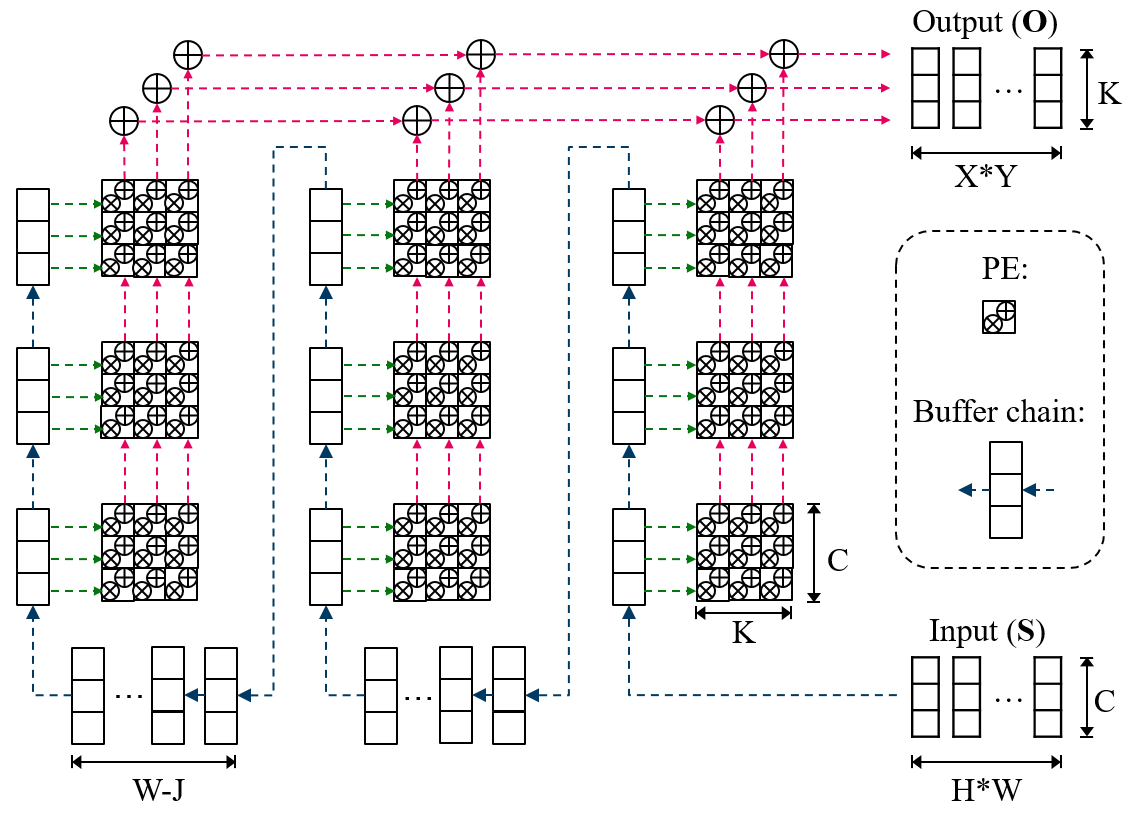}
	\caption{Systolic array architecture for 2D convolution with 3D inputs.}
	\label{fig:systolic_dataflow}
\end{figure}

The buffer chain and the PE array in Fig. \ref{fig:hw_architecture} realize the systolic array that can efficiently process the \textit{convolutional layer}. The computation of a typical convolutional layer is a 2D convolution with 3D inputs, which can be represented by a 6-loop algorithm: 
\begin{algorithmic}
	\State $For(k=0;k<K;k++)\ \ \ \ \ \ \ \ // K = kernel\ num.$
	\State $\ For(x=0;x<X;x++)\ \ \ \ \ \ // X = output\ height$
	\State $\ \ For(y=0;y<Y;y++)\ \ \ \ \ // Y = output\ width$
	\State $\ \ \ For(c=0;c<C;c++)\ \ \ \ \ // C = input\ channel$
	\State $\ \ \ \ For(i=0;i<I;i++)\ \ \ \ \ // I = kernel\ height$
	\State $\ \ \ \ \ For(j=0;j<J;j++)\ \ // J = kernel\ width$
	\State $\ \ \ \ \ \ \bm{O}[k, x, y] += \bm{W}[k, c, i, j] \times \bm{S}[c, x+i, y+j];$
\end{algorithmic}
All these loops can be computed by the proposed systolic array architecture as shown in Fig. \ref{fig:systolic_dataflow}, where the PE array and the buffer chain are properly arranged based on the convolution shape parameters, i.e., $C, H, W, I, J, X,$ and $Y$, where $H$ and $W$ are input height and width. Therefore, the PE array has $CJ$ rows and $KI$ columns, and can be partitioned into $J \times I$ sub-arrays. Each sub-array consists of $C \times K$ PEs arranged in a crossbar structure as shown in Fig. \ref{fig:systolic_dataflow}. Each PE stores a weight value and performs a \textit{multiply-accumulate} (MAC) operation. As to the buffer chain, a total of $(I-1)W+J$ buffers are connected together, and each buffer stores $C$ words of data. Among them, $I \times J$ buffers are used as the input of the PE crossbar arrays, and others are used for delay adjustment. 

\begin{figure}[h]
	\centering
	\includegraphics[width=\linewidth]{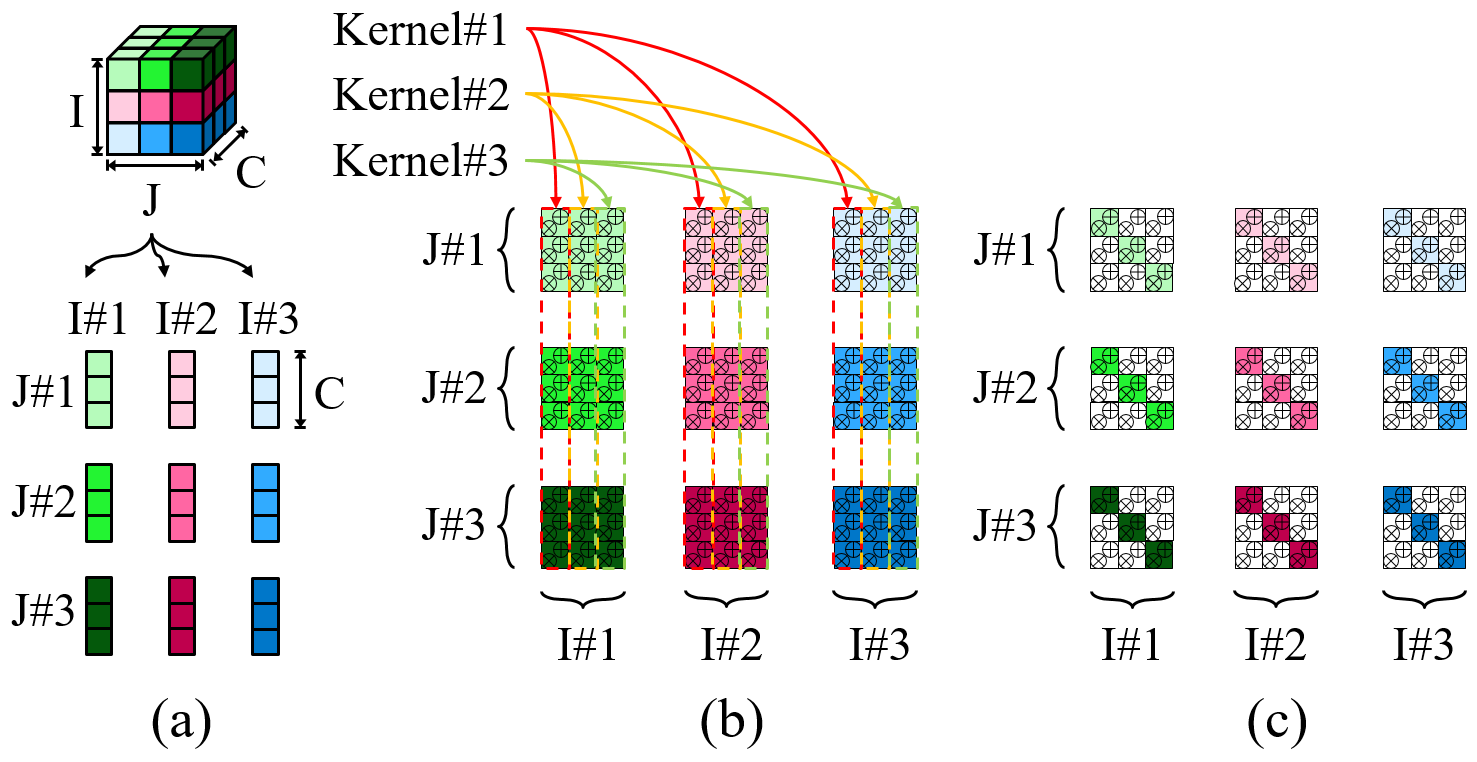}
	\caption{(a) Reshape kernels from 3D to 2D. (b) Map kernels to a PE array. (c) Mapping scheme for depthwise convolution.}
	\label{fig:kernel_map}
\end{figure}

Figure \ref{fig:kernel_map} shows our kernel mapping scheme that maps $K$ 3D-kernels to the PE array. As shown in Fig. \ref{fig:kernel_map}(a), every kernel is reshaped from $(I,J,C)$ to $(I,JC)$, then transposed to $(JC,I)$. These $K$ 2D-kernels are mapped to the PE array as shown in Fig. \ref{fig:kernel_map}(b). 

Figure \ref{fig:systolic_dataflow} shows the data scheduling for the input matrix $\bm{S}$, which is partitioned into $H \times W$ vectors. Each vector contains $C$ data words, which are originally at the same plane of $\bm{S}$, but different channels. One vector enters the buffer chain in every cycle. In the buffer chain, the data follow the arrow direction and go to the next buffer. The buffers next to the PE crossbar arrays are the inputs of those arrays. These data are broadcast and multiplied by the weights stored in PEs. The products are then accumulated across PEs to produce $K$ results of $\bm{O}$. As shown in Fig. \ref{fig:systolic_dataflow}, the output order of $\bm{O}$ is the same as the input order of $\bm{S}$.

While the proposed systolic array is designed for the computation of a convolutional layer, it can also support the \textit{fully-connected (FC) layer, depthwise convolutional layer}, and \textit{average pooling layer}. The FC layer can be seen as a convolutional layer with kernel size 1, i.e., $I=J=1$. Depthwise convolution can be achieved by diagonally mapping the kernels to the PE array as shown in Fig. \ref{fig:kernel_map}(c). Average pooling can be considered as a depthwise convolution with all the weight values set to 1.

This architecture has two features. First, all the data are maximally reused in the systolic array. In other words, once the data flows into the systolic array, it does not flow out unless it is no longer needed. Each data only needs to be accessed from the off-chip memory once. Therefore, the power consumption decreases by reducing off-chip memory access. Second, the output order of $\bm{O}$ is the same as the input order of $\bm{S}$. Therefore, every time an output is generated, it becomes the input of the next systolic array.

\begin{figure}[h]
	\centering
	\includegraphics[width=\linewidth]{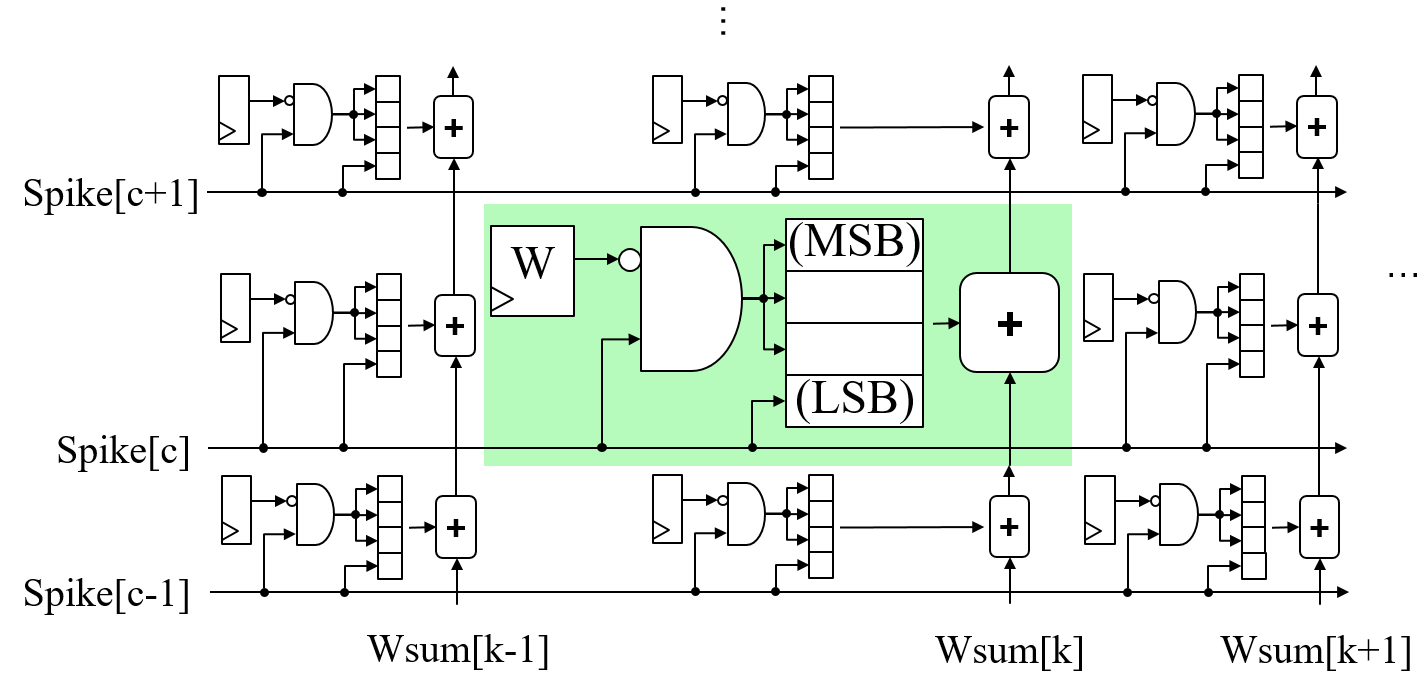}
	\caption{PE crossbar array design for BW-SNN.}
	\label{fig:PEarray_design}
\end{figure}

Fig. \ref{fig:PEarray_design} shows the design of a PE and the construction of the PE crossbar array. In the figure, an AND gate is used for the multiplication of the spike value and the weight value. Because both the spike $\{0,1\}$ and the weight $\{-1,+1\}$ are single bit in a BW-SNN, the possible product belongs to the set $\{-1,0,+1\}$. Assume the weight -1 is stored as 0 in the PE, then the multiplication can be computed by a simple logic equation, i.e., $o=\{\overline{w} \& s, s\}$. As a result, the design of a PE simply consists of an adder, an AND gate, and a flip-flop (FF) storing the inverse of a 1-bit weight.

\subsection{Support for Advanced Network Architectures}
\begin{figure}[h]
	\centering
	\begin{subfigure}[b]{0.48\linewidth}
		\includegraphics[width=\linewidth]{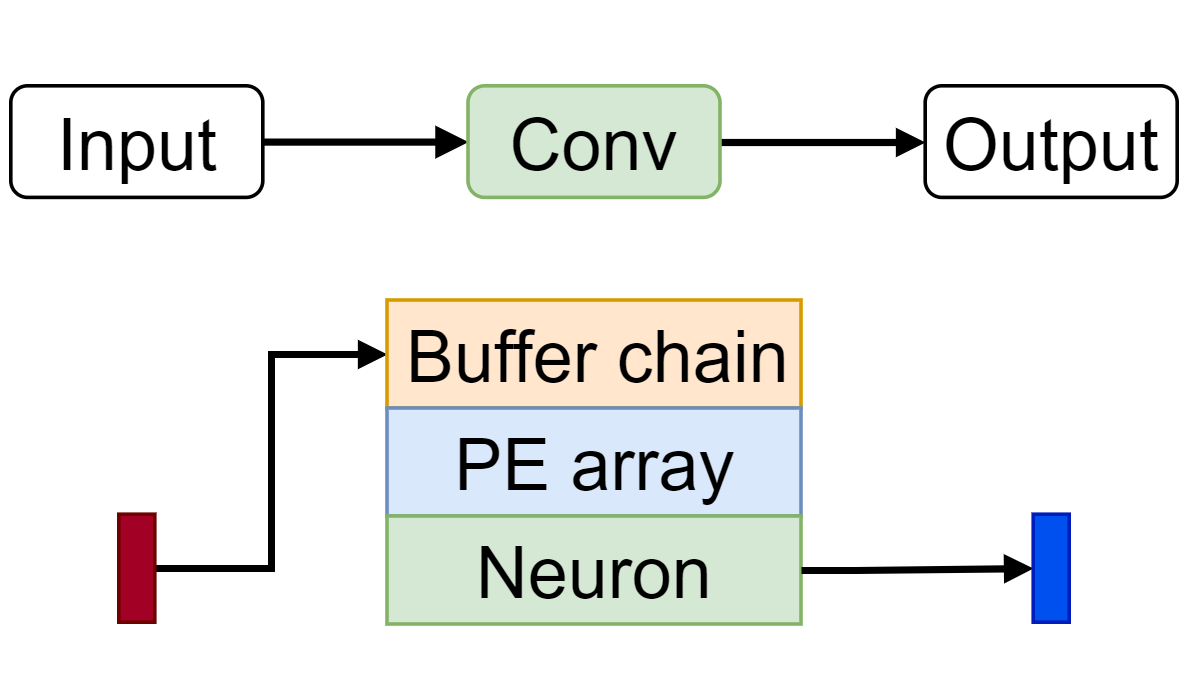}
		\caption{}
		\label{fig:conv_layer}
	\end{subfigure}
	\hfill
	\begin{subfigure}[b]{0.48\linewidth}
		\includegraphics[width=\linewidth]{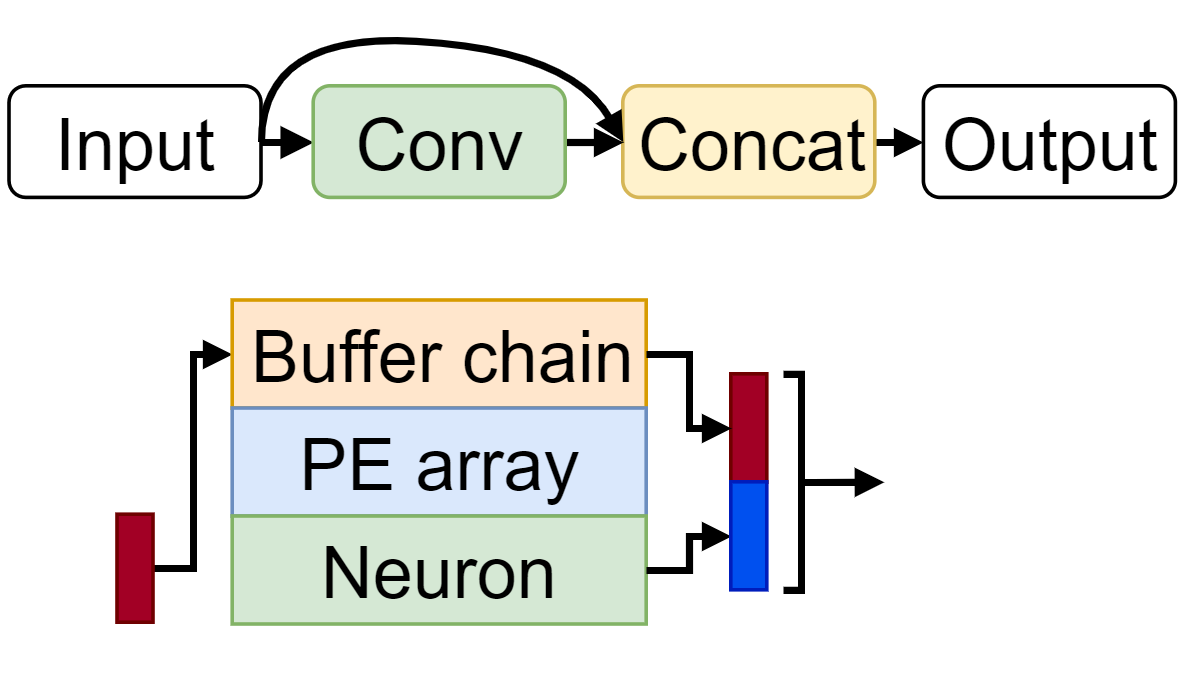}
		\caption{}
		\label{fig:skip1_layer}
	\end{subfigure}
	\\
	\begin{subfigure}[b]{0.48\linewidth}
		\includegraphics[width=\linewidth]{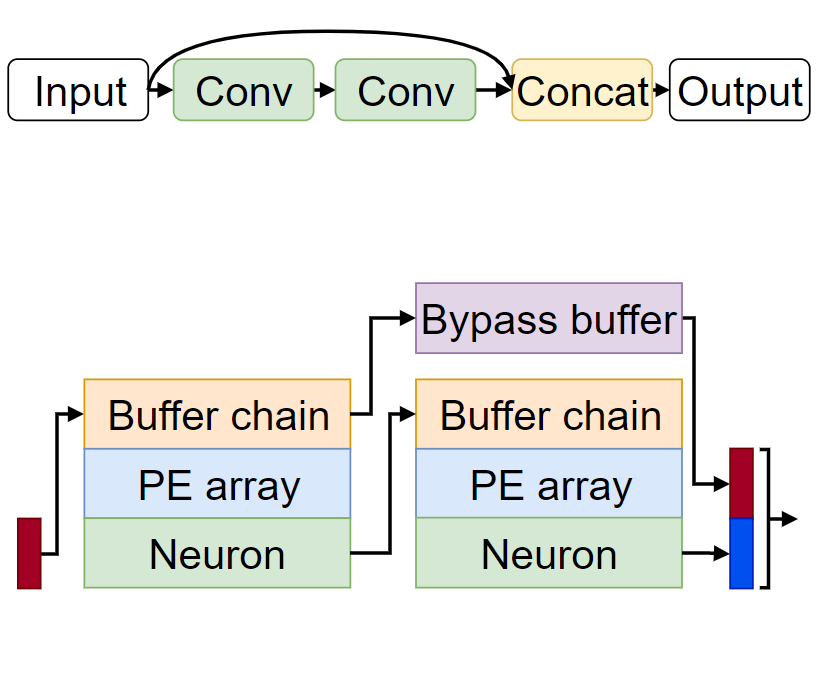}
		\caption{}
		\label{fig:skip2_layer}
	\end{subfigure}
	\hfill
	\begin{subfigure}[b]{0.48\linewidth}
		\includegraphics[width=\linewidth]{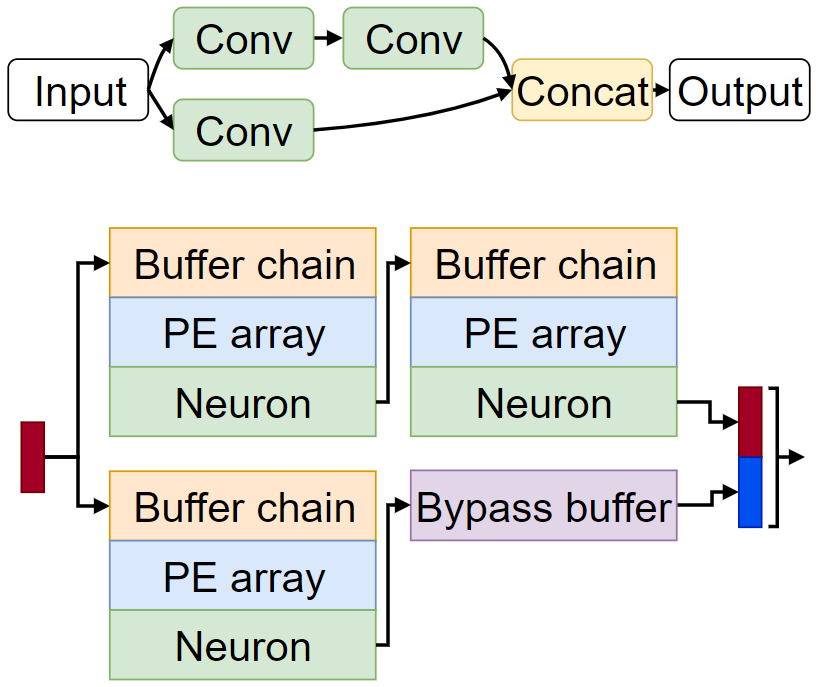}
		\caption{}
		\label{fig:branch_layer}
	\end{subfigure}
	\caption{The design methodologies for different network topologies: (a) typical layer, (b) skip-connect, (c) skip-2-connect, (d) branch.}
	\label{fig:variant_layers}
\end{figure}

Recent neural networks, such as MobileNet \cite{howard2017mobilenets}, VGG \cite{simonyan2014very}, DenseNet \cite{huang2017densely}, and Inception \cite{szegedy2016rethinking}, have their own unique features and network architectures. Fig. \ref{fig:variant_layers} shows the design methodologies to support these varied network topologies. For example, the typical CNN models, such as VGG and MobileNet, can be implemented by stacking the basic layer modules as shown in Fig. \ref{fig:conv_layer}. There are variations for the basic layer module, e.g., convolution, depthwise convolution, fully-connect, and average-pooling, each with a proper systolic array implementation as discussed in Sec. \ref{sec:systolic}.

However, modern neural networks have more complicated topologies, e.g., Figs. \ref{fig:skip1_layer} and \ref{fig:skip2_layer} show the \textit{skip-connection} schemes that are used in DenseNet, where the input of a layer can branch and bypass one or more layers before concatenating with the output of the current or later layer. To support this scheme, the buffer chain in the layer module is used as the input bypass path if the input skips only the current layer, as shown in Fig. \ref{fig:skip1_layer}. If the input skips more than one layer, additional bypass buffers are inserted into the subsequent layer modules to form a longer bypass path, as shown in Fig. \ref{fig:skip2_layer}. Therefore, the skip-connection scheme can be realized by bypass buffers. In Fig. \ref{fig:branch_layer} we show a more complicated branch scheme that is used in the Inception architectures, where the input goes through multiple branches, and the outputs of these reconvergent  branches are concatenated. 

\section{Hardware Implementation Results}
\label{sec:result_hw}

\begin{figure}[h]
	\centering
	\includegraphics[width=\linewidth]{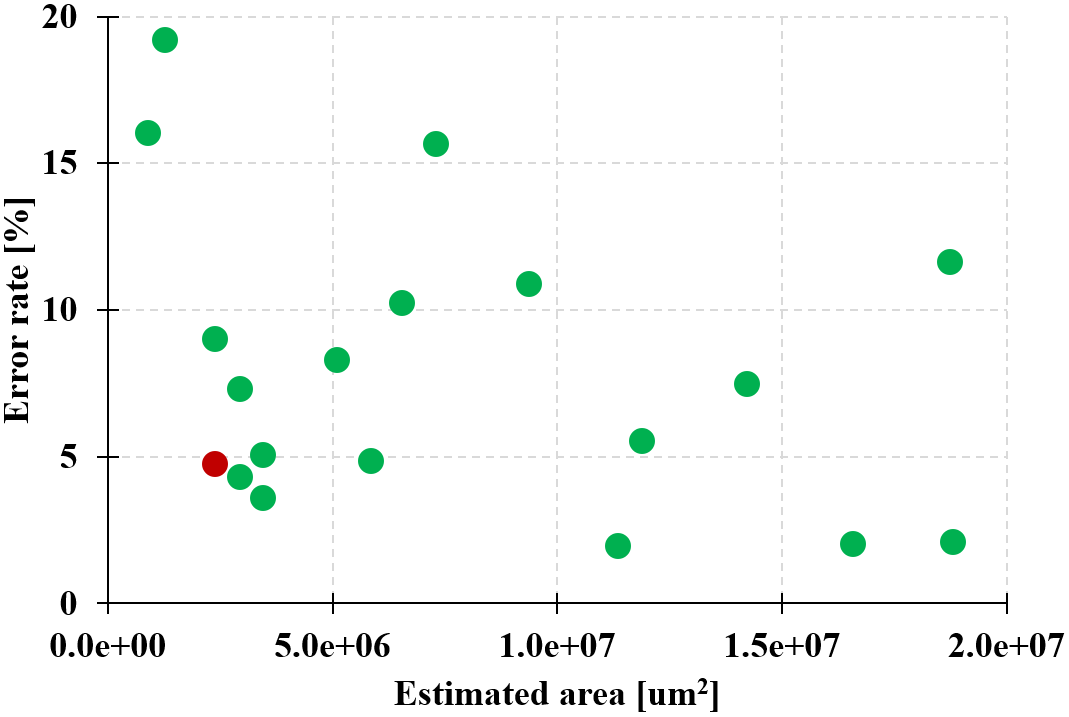}
	\caption{Accuracy-area trade-off for bottled-drink classification.}
	\label{fig:bottle_accuracy_area}
\end{figure}

Our target application is the classification of bottled drinks, with a dataset of 6 classes, i.e., whisky, tequila, cola, lemon juice, orange juice, and pineapple juice. We train a number of BW-SNN models with different network topologies, and estimate their hardware area. We only consider the area of the systolic arrays and local buffers, because they dominate the overall area. For a layer module with the shape parameters of $C, H, W, I, J, X$, and $Y$, the estimated area in um$^2$ of the PE array, the buffer chain, and the local buffer are $210CKIJ, 15C((I-1)W+J)$, and $40KXY$, respectively. Figure \ref{fig:bottle_accuracy_area} shows the accuracy-area tradeoff among different BW-SNN models, where a dot represents a BW-SNN model. We select the model represented by the red dot in the figure, which achieves the lowest area and power with an accuracy of at least 95\%. 

\begin{figure}[h]
	\centering
	\includegraphics[width=\linewidth]{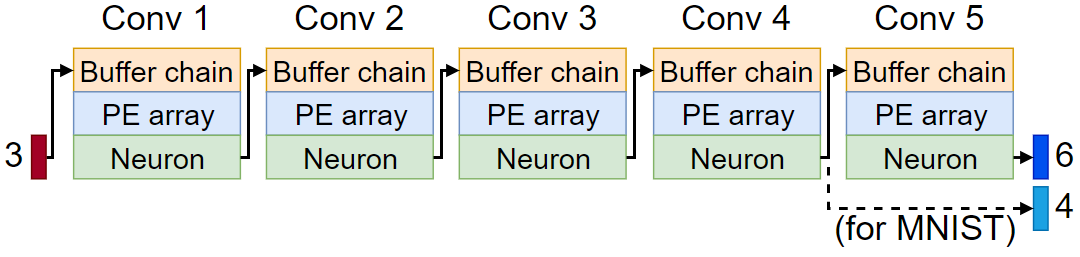}
	\caption{The 5-layer BW-SNN hardware architecture.}
	\label{fig:5conv_bwsnn}
\end{figure}

\begin{figure}[h]
	\centering
	\includegraphics[width=0.7\linewidth]{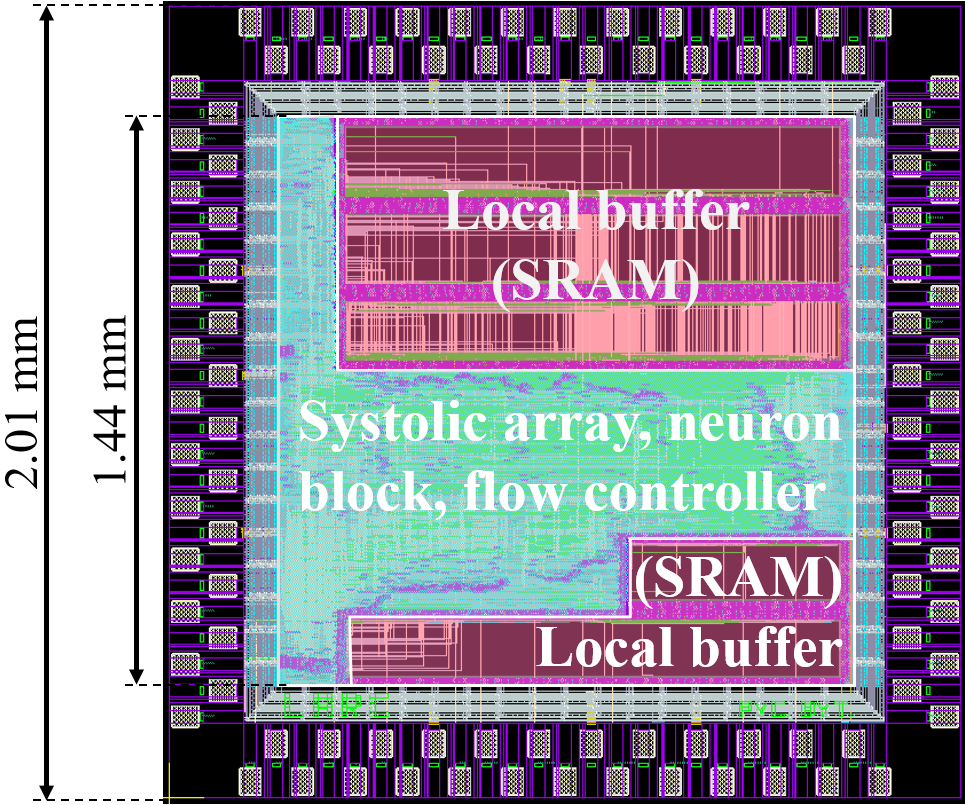}
	\caption{Chip layout.}
	\label{fig:chip_layout}
\end{figure}

Figure \ref{fig:5conv_bwsnn} and Table \ref{tab:hw_nn_arch} show the topology of the selected BW-SNN and the shape parameters for all the layers. This BW-SNN consists of 5 convolutional layers with different numbers of $3 \times 3$ kernels, a stride of 1, and no zero-padding. The chip is implemented in a commercial 90nm CMOS technology with a standard cell-based design flow, and is under manufacturing now. Figure \ref{fig:chip_layout} shows the chip layout, where the equivalent gate count is 225K for the logic part, and the total on-chip SRAM is 12.75K bytes. The operating frequency is 100MHz, and the supply voltages are 1V and 3.3V for the core circuit and the IO pads, respectively, resulting in a power consumption of only 52mW.

\begin{table}[h]
 \begin{center}
  \caption{Layer Shape Parameters for the BW-SNN.}
  \label{tab:hw_nn_arch}
  \begin{tabular}[c]{|c|c c c c c|}
   \hline
   \textbf{Layer} & $\bm{C}$ & $\bm{(H, W)}$ & $\bm{(I, J)}$ & $\bm{K}$ & $\bm{(X, Y)}$ \\
   \hline
   Conv 1 & 3  & 16 & 3 & 16 & 14 \\
   Conv 2 & 16 & 14 & 3 & 16 & 12 \\
   Conv 3 & 16 & 12 & 3 & 16 & 10 \\
   Conv 4 & 16 & 10 & 3 & 16 &  8 \\
   Conv 5 & 16 &  8 & 3 &  6 &  6 \\
   \hline
  \end{tabular}
 \end{center}
\end{table}

For comparison, we evaluate our BW-SNN hardware with the MNIST dataset \cite{lecun-mnisthandwrittendigit-2010}. Although the BW-SNN hardware is designed for bottled-drink classification, we can train new weights for the 5-layer BW-SNN using the MNIST dataset and adjust the weights in the hardware. The post-layout gate-level simulation results for MNIST are summarized in Tab. \ref{tab:hw_mnist}, where power consumption results are obtained from a commercial power simulator with data switching activity information. With a latency of 0.5ms, the hardware can achieve the maximum accuracy of 98.73\%. The energy and latency can be reduced by 5.7$\times$ with an acceptable accuracy drop of 0.72\%. These latency results are sufficient for real-time application.

\begin{figure}[h]
	\centering
	\includegraphics[width=\linewidth]{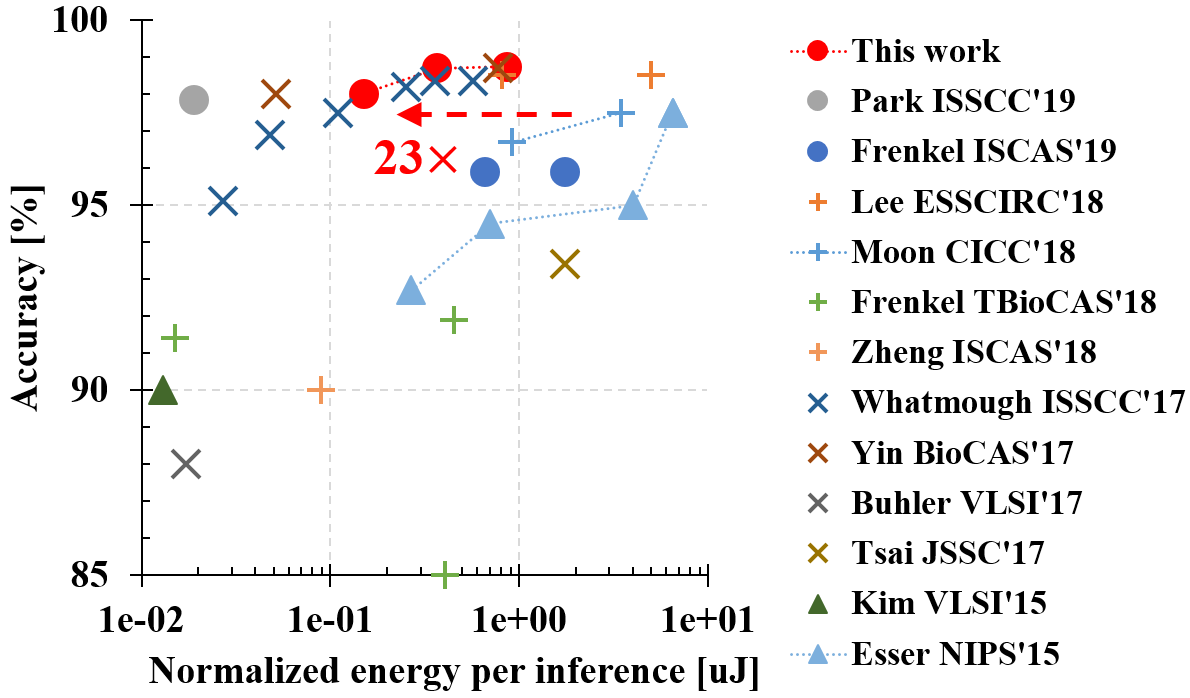}
	\caption{MNIST accuracy and energy comparison with previous works \cite{frenkel201965, lee2018wide, moons2018binareye, frenkel20180, zheng2018low, whatmough201714, park20197, yin2017algorithm, buhler20173, tsai201741, kim2015640m, esser2015backpropagation}. Energy values are normalized to the 28nm node \cite{Rabaey:1996:DIC:227375}, i.e., scaled with $(28/\text{current\_node})^3$.}
	\label{fig:acc_energy_area}
\end{figure}

Figure \ref{fig:acc_energy_area} shows the comparison of accuracy vs. energy among this work and some previous ones. Our 5-CONV BW-SNN design achieves the highest accuracy of 98.73\% under an energy constraint of 10uJ/inf. Although \cite{park20197, yin2017algorithm, whatmough201714} achieve similar accuracy and lower energy as compared with our design, they process only FC layers that require less computation than CONV layers, and may limit its capability of handing real-world applications. Therefore, if we only compare the designs supporting CONV layers, i.e., our design, BinarEye \cite{moons2018binareye}, and TrueNorth \cite{esser2015backpropagation}, then our design reduces the energy by 23$\times$ and achieves a higher accuracy of 98.01\% as compared with 97.5\% by BinarEye. Note that our energy value is based on post-layout simulation, while others are based on chip measurement results. As for the area, if normalized to the 28nm node \cite{Rabaey:1996:DIC:227375} and compared with \cite{frenkel201965}, our 5-CONV BW-SNN reduces the area by 2.6$\times$ and achieves a higher accuracy of 98.73\%. Compared with a 9-CONV DNN design \cite{moons2018binareye}, our design reduces the area by 7$\times$ with only 0.12\% accuracy loss. 

\begin{table}[h]
 \begin{center}
  \caption{BW-SNN Post-Layout Simulation Results.}
  \label{tab:hw_mnist}
  \begin{tabular}[c]{|c|c|c|c|}
   \hline
   Num. of    & Latency per    & Energy per     & MNIST \\
   time steps & inference (ms) & inference (uJ) & accuracy (\%) \\
   \hline
   37 & 0.095 & 4.991 & 98.01 \\
   \hline
   90 & 0.231 & 12.22 & 98.7 \\
   \hline
   212 & 0.543 & 28.719 & 98.73 \\
   \hline
  \end{tabular}
 \end{center}
\end{table}

\section{Conclusion}
\label{sec:conclusion}

In this paper, we propose a hardware architecture for low-cost, low-power, and real-time object classification. The proposed BW-SNN architecture is designed to maximize data reuse and hardware utilization. No off-chip memory bandwidth is required thanks to the proposed novel systolic array that allows efficient data flow in the hardware. A 5-CONV BW-SNN hardware was implemented and being fabricated in a commercial 90nm CMOS technology, demonstrating a real-time, low-power, and low-cost bottled-drink classifier that achieves a power consumption of 52mW, with a die size of only 2.07mm$^2$. Furthermore, extensive simulation on the MNIST benchmark shows that, as compared with previous hardware designs, our design shows 7$\times$ and 23$\times$ reductions on area and energy, respectively, while achieving higher accuracy.

\section*{Acknowledgement}
This work was supported in part by the Ministry of Science and Technology under Grants 107-2218-E-007-037 and 108-2218-E-007-026, the Industrial Technology Research Institute under Grant B5-10703-HQ-01, and Taiwan Semiconductor Research Institute (TSRI) on chip fabrication.

\bibliography{ref}

\end{document}